\documentclass{nature}

\usepackage{graphicx,color,hyperref,amsmath,amssymb}

\newcommand{\onlinecite}[1]{\hspace{-1 ex} \nocite{#1}\citenum{#1}} 

\title{Quantum Monte Carlo study of the phase diagram of solid
  molecular hydrogen at extreme pressures}

\author{N.\ D.\ Drummond$^1$, Bartomeu Monserrat$^2$,
  Jonathan H.\ Lloyd-Williams$^2$, P.\ L\'opez\ R\'{\i}os$^2$, Chris
  J.\ Pickard$^3$, \& R.\ J.\ Needs$^2$}

\begin{document}

\maketitle

\begin{affiliations}
\item Department of Physics, Lancaster University, Lancaster LA1 4YB, UK
\item Theory of Condensed Matter Group, Cavendish Laboratory, University of
  Cambridge, J.\ J.\ Thomson Avenue, Cambridge CB3 0HE, UK
\item Department of Physics \& Astronomy, University College London, Gower
  Street, London WC1E~6BT, UK
\end{affiliations}

\begin{abstract}
  Establishing the phase diagram of hydrogen is a major challenge for
  experimental and theoretical physics. Experiment alone cannot establish the
  atomic structure of solid hydrogen at high pressure, because hydrogen
  scatters X-rays only weakly.  Instead our understanding of the atomic
  structure is largely based on density functional theory (DFT)\@.  By
  comparing Raman spectra for low-energy structures found in DFT searches with
  experimental spectra, candidate atomic structures have been identified for
  each experimentally observed phase.  Unfortunately, DFT predicts a metallic
  structure to be energetically favoured at a broad range of pressures up to
  400 GPa, where it is known experimentally that hydrogen is nonmetallic.
  Here we show that more advanced theoretical methods (diffusion quantum Monte
  Carlo calculations) find the metallic structure to be uncompetitive, and
  predict a phase diagram in reasonable agreement with experiment.  This
  greatly strengthens the claim that the candidate atomic structures
  accurately model the experimentally observed phases.
\end{abstract}

Hydrogen (H) is the simplest and most abundant of all elements and yet it 
displays amazing richness in its phase behaviour\cite{Mao_H_review,
McMahon_H_review_2012}: it is observed to form a quantum crystalline state and 
orientationally ordered molecular phases, and it has been predicted to exhibit 
a liquid-metal phase at high pressures and low temperatures\cite{Deemyad_H_melting_2008,Bonev_quantum_fluid_metallic_hydrogen_2004, Chen_PIMD_2013}, metallic
superfluid and superconducting superfluid
states\cite{Babaev_superconducting_superfluid_2004, Babaev_metallic_superfluid_2005},
and high-temperature
superconductivity\cite{Ashcroft_1968_high_Tc_superconductivity_H_1968,
  Cudazzo_superconducting_H_DFT_2008, McMahon_superconductivity_H}.
Several crystalline phases of solid molecular H have been observed in
diamond anvil cell experiments carried out at pressures up to over 300 GPa\cite{Loubeyre_320_GPa_2002,Eremets_conductive_H_2011,
Howie_molecular_atomic_H_2012,Howie_phase_IV_2012,Zha_H_360GPa_2012,
Loubeyre_H_290GPa_2013,Zha_H_IV_2013,Goncharov_H_2013,Zha_Raman_325GPa_2014}.
The low-pressure phase I, which is a hexagonal close-packed structure formed of
freely rotating molecules, transforms to a broken-symmetry phase II, in which 
the molecular rotations are restricted, at low temperatures\cite{Mao_H_review,
McMahon_H_review_2012}. 
The transition pressure decreases strongly 
with isotopic mass\cite{Mao_H_review,Cui_D2_1994,Moshary_HD_1993,Mazin_1997,
Goncharov_H_D_200GPa_2011} and also depends on the total spin of the molecules\cite{Mao_H_review,Mazin_1997}. 
As the pressure is increased at low temperatures, there is a further transition 
from phase II to a phase III at about 160 GPa, with the transition pressure for
deuterium (D) exceeding that for H by about 12 GPa\cite{Goncharov_H_D_200GPa_2011}.
Experimental studies have also demonstrated the existence of a phase IV at 
temperatures above a few hundred K and pressures above 220 GPa\cite{Eremets_conductive_H_2011,Howie_molecular_atomic_H_2012,
Howie_phase_IV_2012,Zha_H_IV_2013,Loubeyre_H_290GPa_2013}.
Some constraints on the structures of the observed phases have been obtained 
from X-ray diffraction experiments\cite{Goncharenko_II_2005,
Akahama_x-ray_H_2010}, but the low X-ray scattering cross section of H and the 
small sample sizes available limit the possible resolution. 
Infrared (IR) and particularly Raman spectroscopic measurements have yielded 
valuable information about the vibrational modes of H at high pressures\cite{Loubeyre_320_GPa_2002,Eremets_conductive_H_2011,
Howie_molecular_atomic_H_2012,Howie_phase_IV_2012,Zha_H_360GPa_2012,
Loubeyre_H_290GPa_2013,Zha_H_IV_2013,Goncharov_H_2013,Zha_Raman_325GPa_2014,
Cui_D2_1994,Moshary_HD_1993,Mazin_1997,Goncharov_H_D_200GPa_2011,
Goncharenko_II_2005}, but the available experimental data are insufficient to 
determine the structures of phases II, III, and IV\@.

\begin{figure}
  \includegraphics[clip,width=0.3\textwidth]{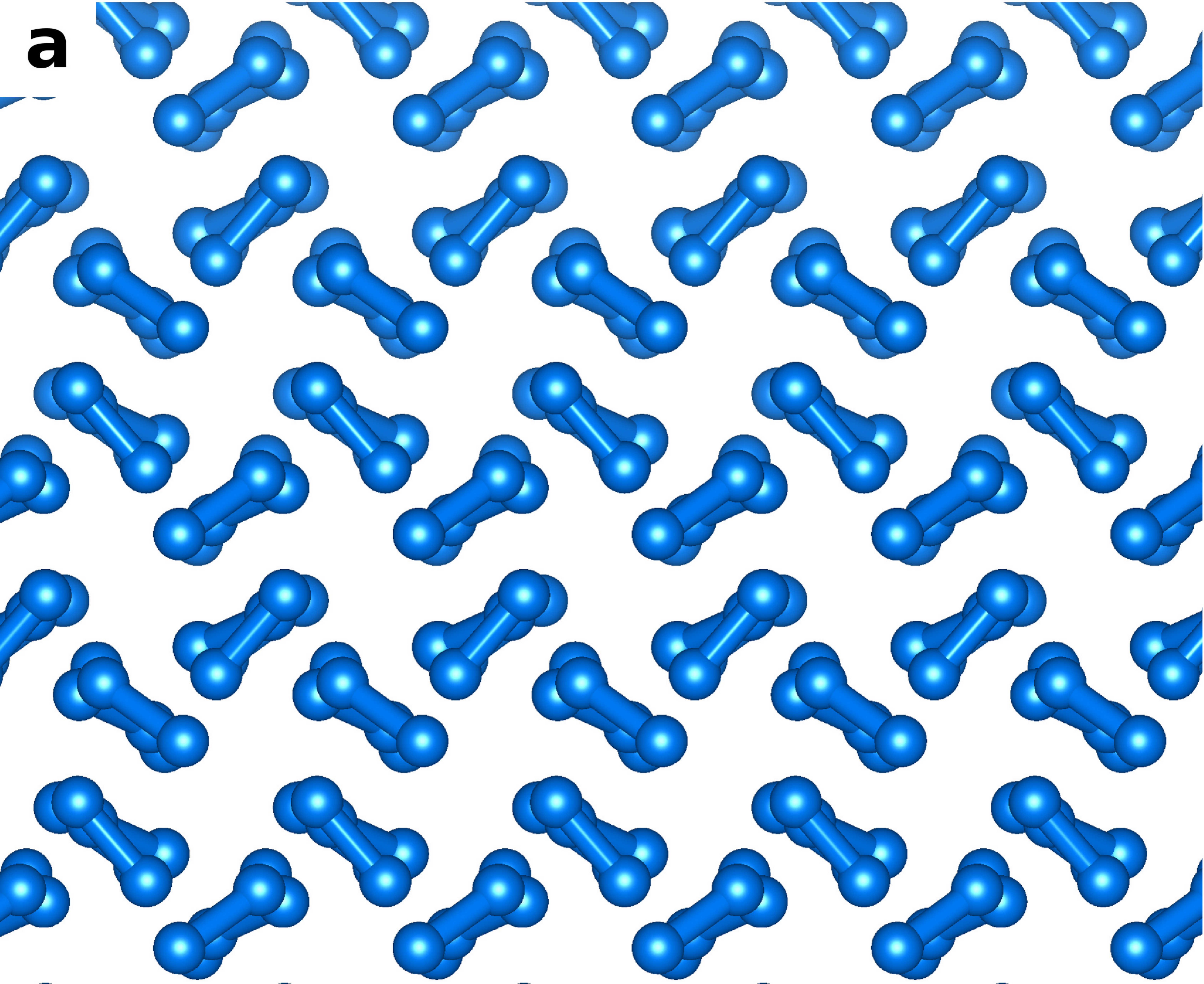}
  \includegraphics[clip,width=0.3\textwidth]{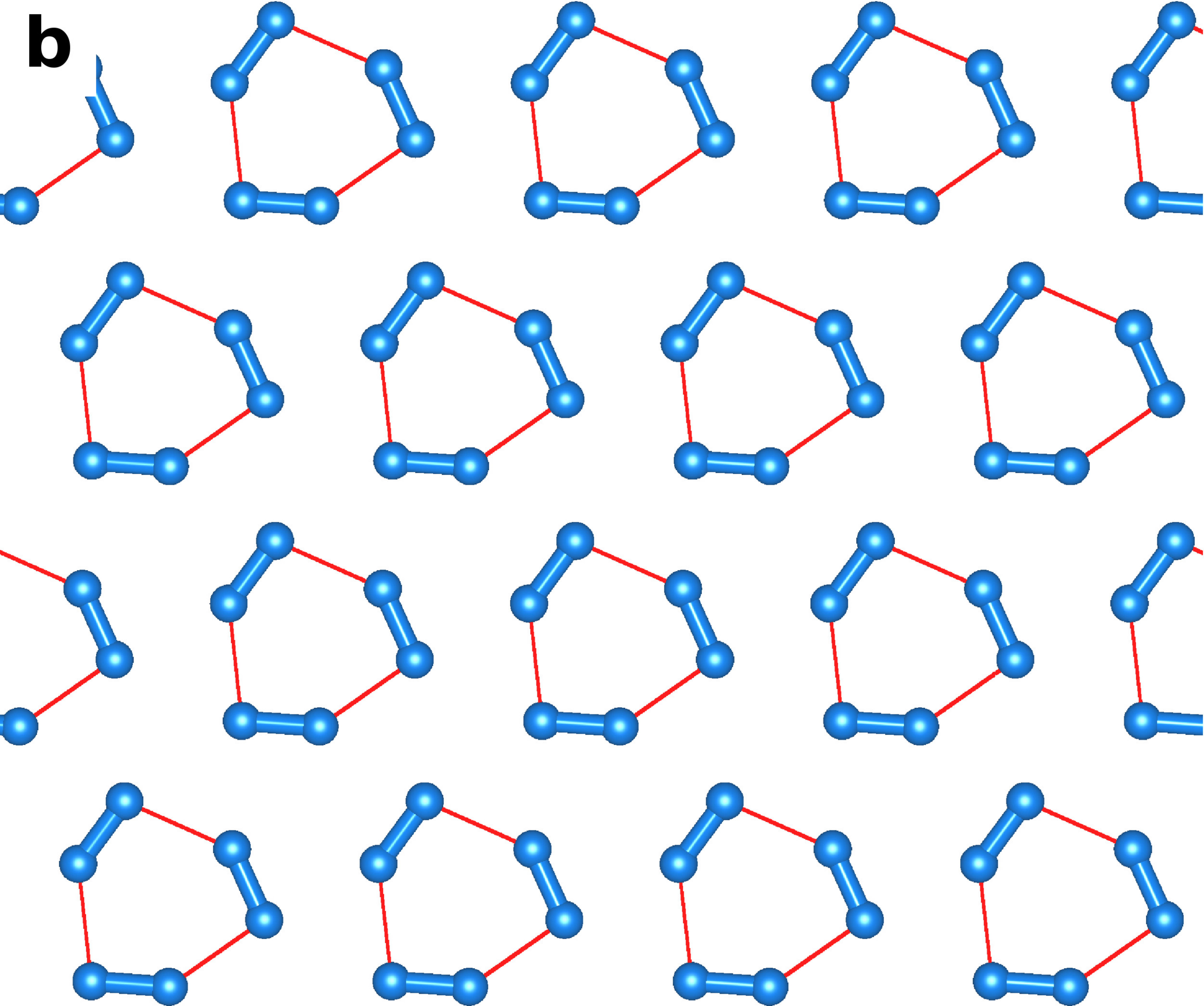}
  \includegraphics[clip,width=0.3\textwidth]{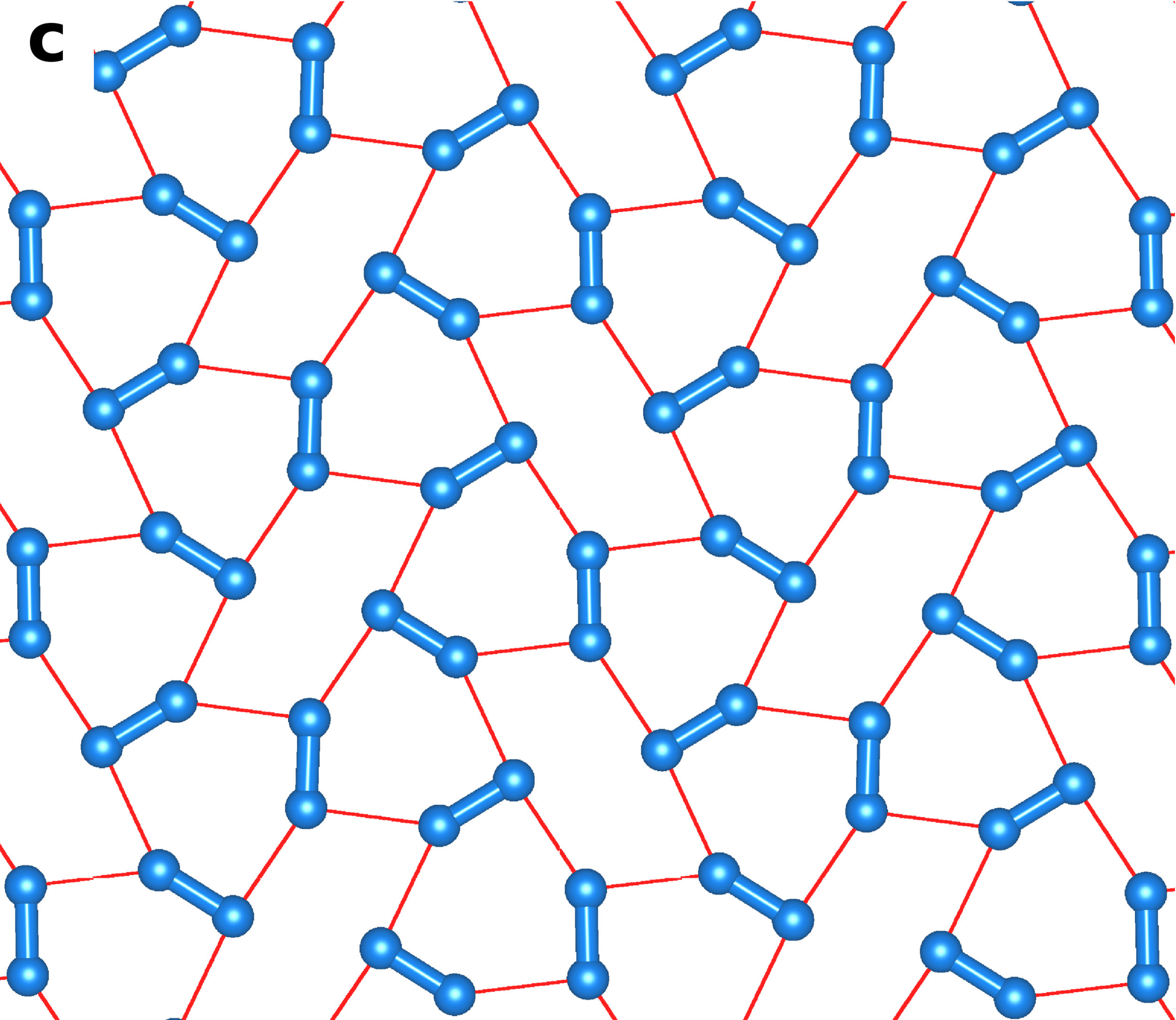}
  \includegraphics[clip,width=0.3\textwidth]{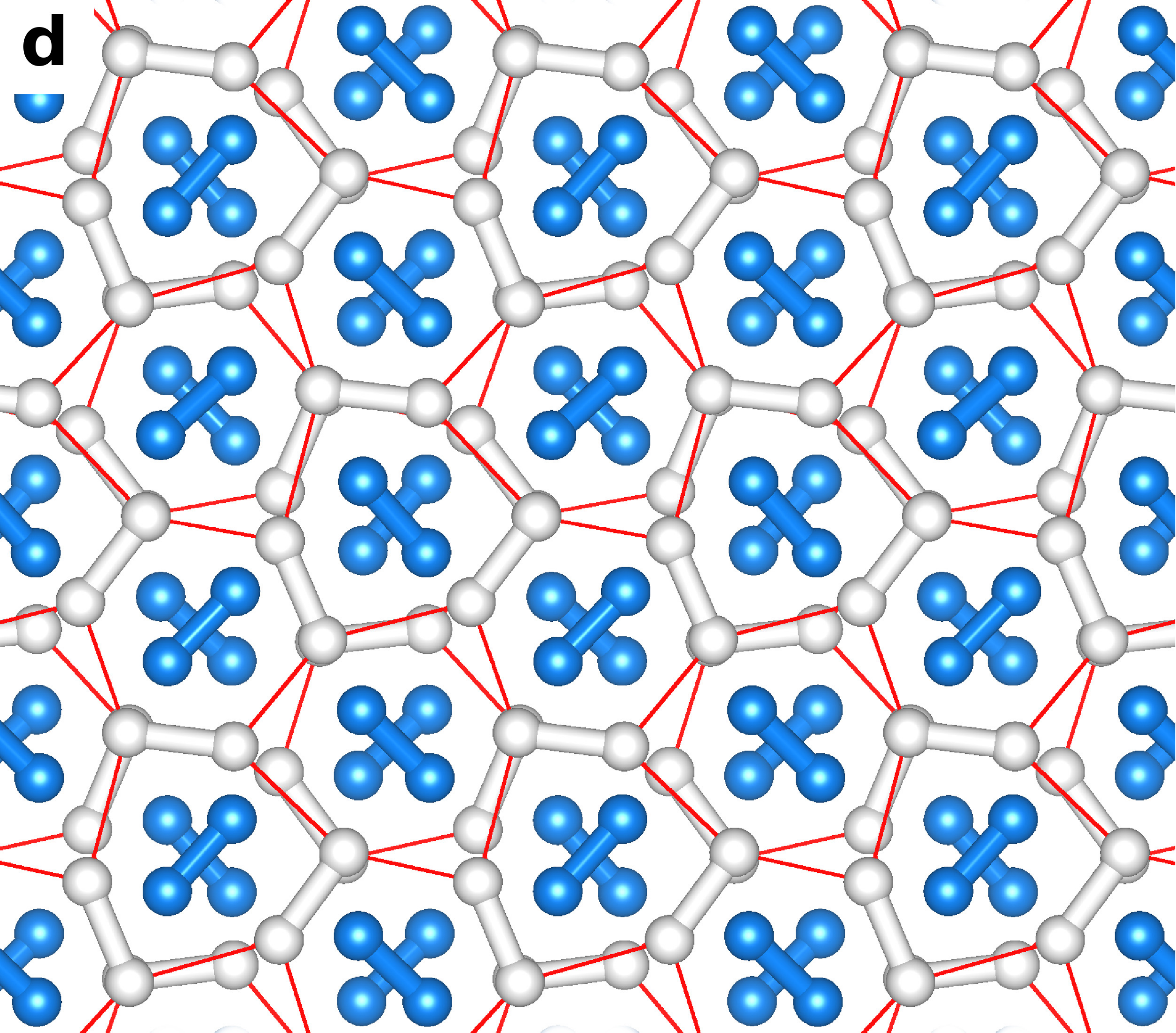}
  \includegraphics[clip,width=0.3\textwidth]{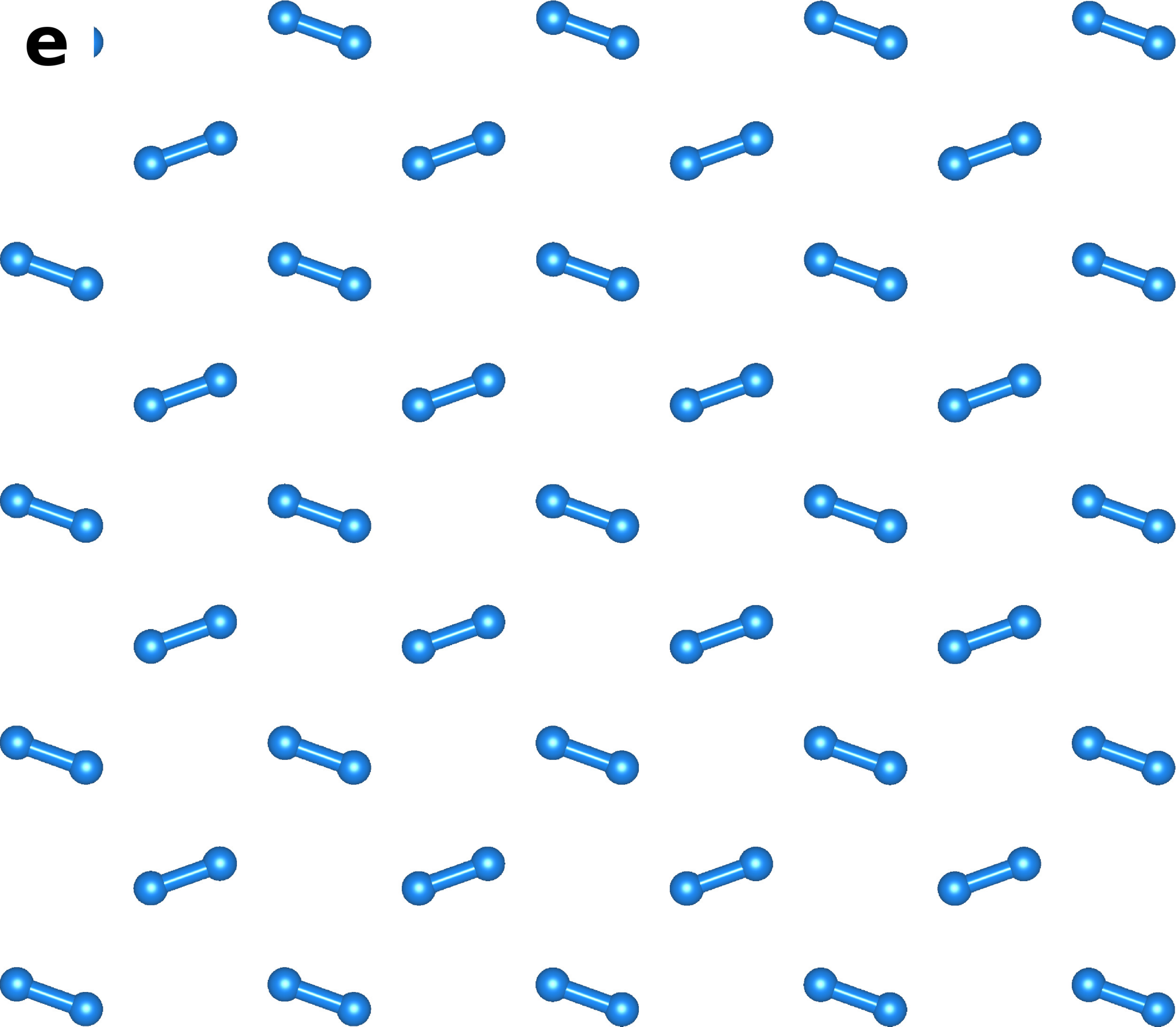}
  \caption{\textbf{Atomic structures of the five H phases considered in this 
    work.}
    (\textbf{a}) \textit{P}2$_{\text{1}}$/\textit{c}-24, 
    (\textbf{b}) \textit{C}2/\textit{c}-24, (\textbf{c}) \textit{Cmca}-12, 
    (\textbf{d}) \textit{Pc}-48, and (\textbf{e}) \textit{Cmca}-4.
    The blue dumbbells indicate short bonds between atoms (\textless 0.8 \AA)\@.
    The white dumbbells indicate long bonds between atoms (\textless 0.9 \AA)\@.
    The red lines indicate close contacts between atoms (\textless 1.2 \AA) 
    in the layered structures. 
    \textit{P}2$_{\text{1}}$/\textit{c}-24 consists of molecules arranged on a 
    distorted hexagonal-close-packed lattice.
    \textit{C}2/\textit{c}-24, \textit{Cmca}-12, and \textit{Cmca}-4 consist of 
    layers of molecules whose bonds lie within the planes of the layers, 
    forming distorted hexagonal patterns, and we show top-down views of single 
    layers.
    \textit{Pc}-48 consists of alternate layers of isolated strongly bonded 
    molecules and weakly bonded graphene-like sheets, and we show a top-down
    view of four layers.
    The structures are shown at a common DFT-PBE pressure (250 GPa).
      \label{fig:picture_structures}}
\end{figure}

Candidate structures for phases II, III, and IV have been suggested by 
structure searches based on density functional theory (DFT)\cite{Pickard_H_2007,
Pickard_some_H_structures_2009,Pickard_IV_2012,Pickard_IV_2012_Erratum,
McMahon_H_structures_2011,Liu_H_2012,Liu_molecular_atomic_2012}, although it 
should be emphasised that none of these structures has been identified as being 
unambiguously correct.
The candidate structures for phase II consist of packings of molecules
with bond lengths almost identical to the zero-pressure value\cite{Pickard_H_2007,Pickard_some_H_structures_2009}. 
We have modelled phase II using a molecular structure of 
\textit{P}2$_{\text{1}}$/\textit{c} symmetry with 24 atoms in the primitive unit 
cell, which we refer to as \textit{P}2$_{\text{1}}$/\textit{c}-24; see Fig.\ 
\ref{fig:picture_structures}(a). 
(We adopt the convention of labelling structures by their symmetry followed by 
the number of atoms per primitive cell.) 
\textit{P}2$_{\text{1}}$/\textit{c}-24 is the most stable structure
found to date in static-lattice DFT within the pressure range appropriate for
phase II, and its vibrational
characteristics are also compatible with those of phase II\@.
We model phase III using a \textit{C}2/\textit{c}-24 structure consisting of 
layers of molecules whose bonds lie within the planes of the layers, forming a 
distorted hexagonal pattern\cite{Pickard_H_2007}; see Fig.\ 
\ref{fig:picture_structures}(b).
This very stable structure can account for the high IR activity of phase III\cite{Pickard_H_2007}.
We also consider a molecular \textit{Cmca}-12 structure\cite{Pickard_H_2007}, 
which is similar to \textit{C}2/\textit{c}-24, but slightly denser; see Fig.\ 
\ref{fig:picture_structures}(c).
We model phase IV by a \textit{Pc}-48 structure\cite{Pickard_IV_2012,
  Pickard_IV_2012_Erratum}, shown in Fig.\
\ref{fig:picture_structures}(d), which consists of alternate layers of
strongly bonded molecules and weakly bonded graphene-like sheets.
This type of structure was predicted by Pickard and Needs\cite{Pickard_H_2007}.
\textit{Pc}-48 can account for the occurrence of stiff and soft
vibronic modes in phase IV, and its stabilisation by temperature.
Finally, we consider the \textit{Cmca}-4 structure\cite{Johnson_2000}, which 
has weaker molecular bonds than \textit{C}2/\textit{c}-24 and \textit{Cmca}-12, 
and is shown in Fig.\ \ref{fig:picture_structures}(e).
The main goals of our present work are to obtain accurate theoretical results 
for the relative stabilities of the \textit{P}2$_{\text{1}}$/\textit{c}-24, 
\textit{C}2/\textit{c}-24, \textit{Cmca}-12, \textit{Pc}-48, and 
\textit{Cmca}-4 structures of H at pressures of 100--400 GPa and temperatures 
of 0--500 K, and to use these data to construct a temperature-pressure phase 
diagram of H\@.
We have not considered phase I in our calculations, which is stable at
low pressures, because an accurate description of this phase would
require a full quantum treatment of the proton spin.  Instead we focus
our attention on the phase behaviour at higher pressures, where the
candidate structures are such that the nuclei are highly localised and hence the
motion of the protons is likely to be well-described by collective
bosonic vibrational modes.

Useful theoretical descriptions of solid H require very accurate calculations
with an energy resolution of a few meV per atom. 
Various studies have shown that DFT currently cannot provide such accuracy for
H structures, as evidenced by the disagreement of results obtained with 
different exchange-correlation functionals and the fact that DFT predicts H to 
be metallic at pressures above $\sim$ 300 GPa, in contradiction with experiment\cite{Pickard_H_2007,Pickard_IV_2012,Pickard_IV_2012_Erratum,
Azadi_DFT_for_H_2013,Morales_H_2013,Clay_benchmarking_DFT_2014}.
We have instead used the diffusion quantum Monte Carlo (DMC) method\cite{Foulkes_2001} to calculate static-lattice energy-volume relations for the different H 
phases.
DMC is generally regarded as the most accurate first-principles method 
available for carrying out such studies\cite{Natoli_1995,Azadi_QMC_H_2013,Azadi_H_QMC_VSCF_2014}.
Furthermore, the low mass of the H atom means that a full treatment of quantum 
nuclear vibrational motion, including anharmonic effects\cite{Morales_H_2013,
Azadi_H_QMC_VSCF_2014}, is crucial for an accurate description of the 
energetics. 
We have therefore used a DFT-based vibrational self-consistent field
approach\cite{Monserrat_VSCF_2013} to calculate anharmonic vibrational
energies.
We find that the use of DMC (and to a lesser extent the treatment of phonon
anharmonicity) renders the metallic \textit{Cmca}-4 structure that is favoured
in DFT energetically uncompetitive, leaving us with a phase diagram in
reasonable quantitative agreement with experiment.

\section*{Results}

\begin{figure}
   \includegraphics[width=0.3\textwidth]{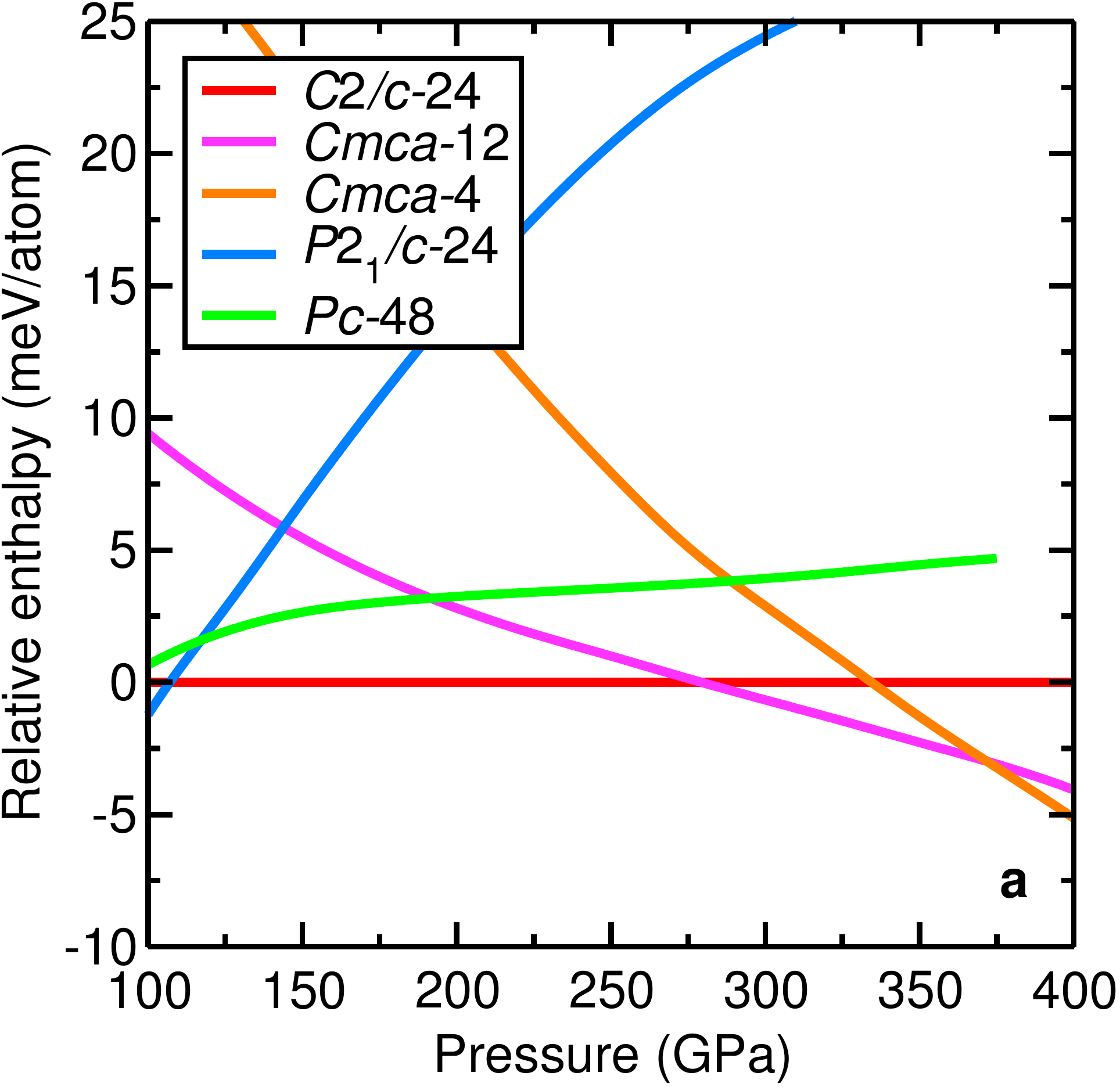}
   \includegraphics[width=0.3\textwidth]{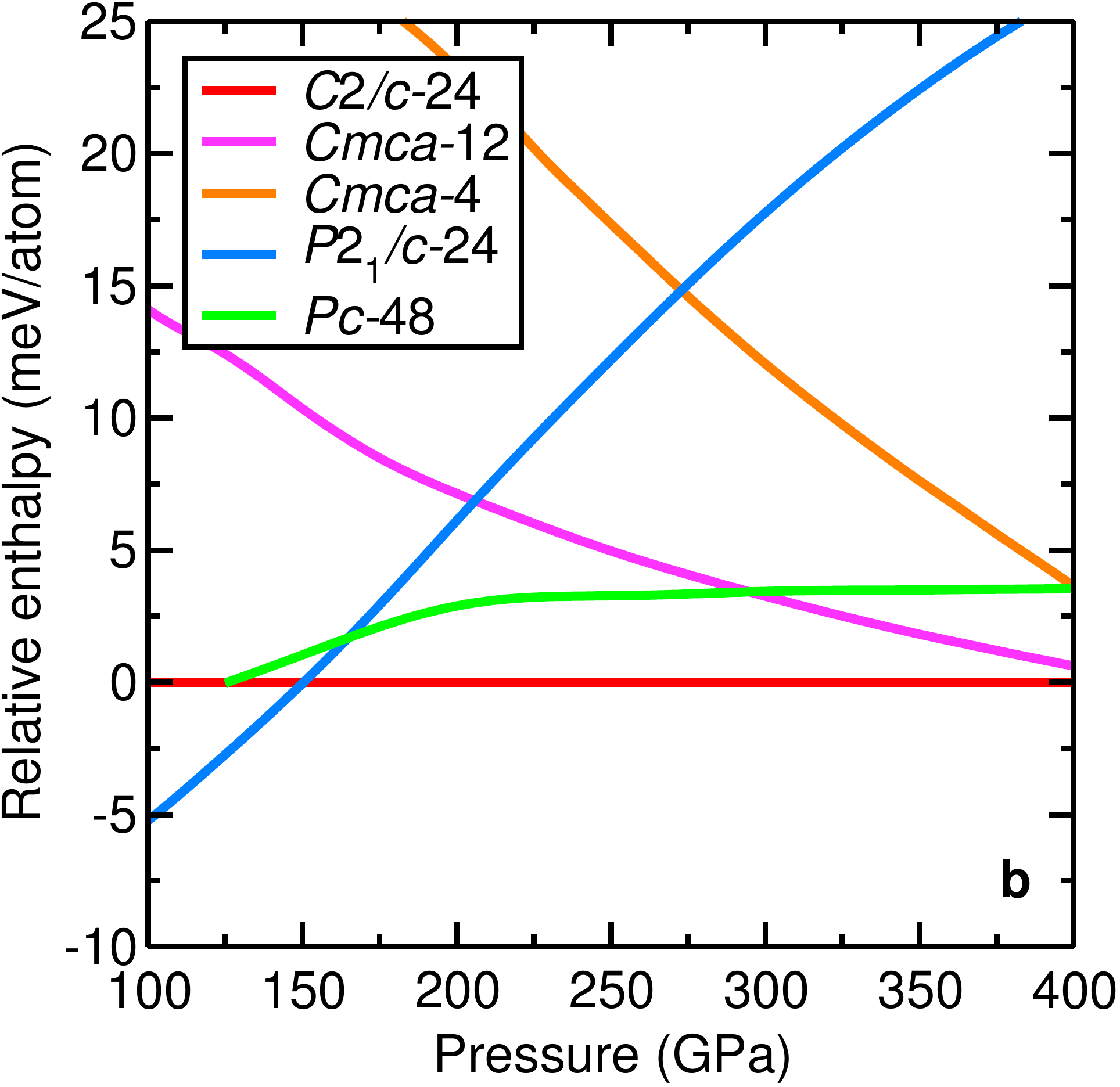}
   \includegraphics[width=0.3\textwidth]{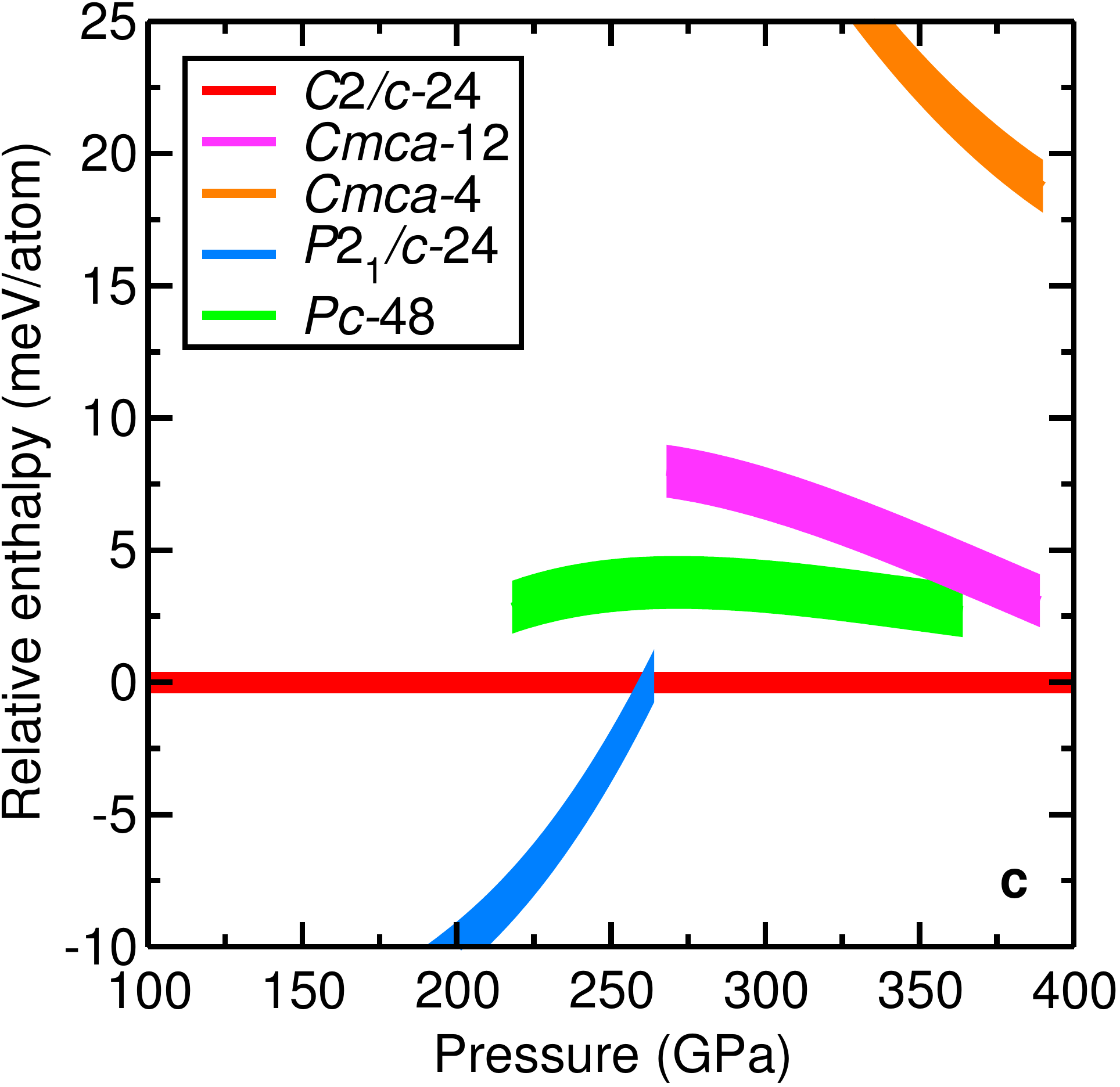}
  \caption{\textbf{DFT and DMC static-lattice enthalpy-pressure relations for 
    the different H structures relative to \textit{C}2/\textit{c}-24.} 
    (\textbf{a}) DFT-PBE, (\textbf{b}) DFT-BLYP, and (\textbf{c}) DMC\@.
    The relative DFT enthalpies are converged to better than 0.1 meV per atom.
    The widths of the DMC lines indicate the estimated uncertainties in the 
    enthalpies due to single-particle finite-size errors, which are greater
    than the uncertainties due to random sampling in the Monte Carlo
    algorithm, as explained in Supplementary Note 2.
  \label{fig:DMC_vs_DFT}}
\end{figure}

\paragraph{Relative enthalpies}
Figure \ref{fig:DMC_vs_DFT} shows the static-lattice enthalpies of the
structures relative to \textit{C}2/\textit{c}-24. 
In Figs.\ \ref{fig:DMC_vs_DFT}(a) and \ref{fig:DMC_vs_DFT}(b) we report DFT 
enthalpies calculated using the Perdew-Burke-Ernzerhof (PBE)\cite{PBE_1996} and 
Becke-Lee-Yang-Parr (BLYP) density functionals\cite{LYP_1988,Becke_1988}.
The relative DFT enthalpies are converged to better than 0.1 meV per atom with 
respect to ${\mathbf{k}}$-point sampling and plane wave cutoff energy. 
The difference between the DFT-PBE and DFT-BLYP relative enthalpies arises 
chiefly from the energetics and not from the slightly different structures 
obtained from geometry optimisation calculations performed at fixed external
pressures using the two different functionals: see Supplementary Note 1
and the accompanying Supplementary Fig.\ 1.
In Fig.\ \ref{fig:DMC_vs_DFT}(c) we report DMC enthalpies, which were obtained 
by fitting polynomials to the extrapolated infinite-system-size DMC energies as 
a function of volume, and differentiating the polynomials to obtain pressures. 
The structures used for the DMC calculations were obtained from DFT-PBE
geometry optimisation calculations.
We truncate the DMC enthalpy curves at the highest and lowest pressures at 
which we have performed calculations.

The use of the DMC method has significant consequences for the static-lattice 
relative enthalpies of the candidate structures.
Compared with both DFT-PBE and DFT-BLYP, \textit{Cmca}-4 and \textit{Cmca}-12 
are destabilised with respect to \textit{C}2/\textit{c}-24, whereas 
\textit{P}2$_{\text{1}}$/\textit{c}-24 is stabilised with respect to it,
but in each case the DFT-BLYP results are closer to the DMC enthalpies, as also 
found in Ref.\ \onlinecite{Clay_benchmarking_DFT_2014}.
For \textit{Cmca}-4 and \textit{P}2$_{\text{1}}$/\textit{c}-24 the difference 
between the DMC and the DFT-BLYP results is greater than the difference between 
the DFT-BLYP and DFT-PBE results, while for \textit{Cmca}-12 these differences 
are of similar size.
Although DFT-BLYP happens to be relatively accurate in the pressure range
of interest, it is clear that DFT is unable to provide a consistent,
quantitative description of the relative enthalpies of the phases of H\@.

\begin{figure}
\begin{tabular}{ll}
   \includegraphics[width=0.3\textwidth]{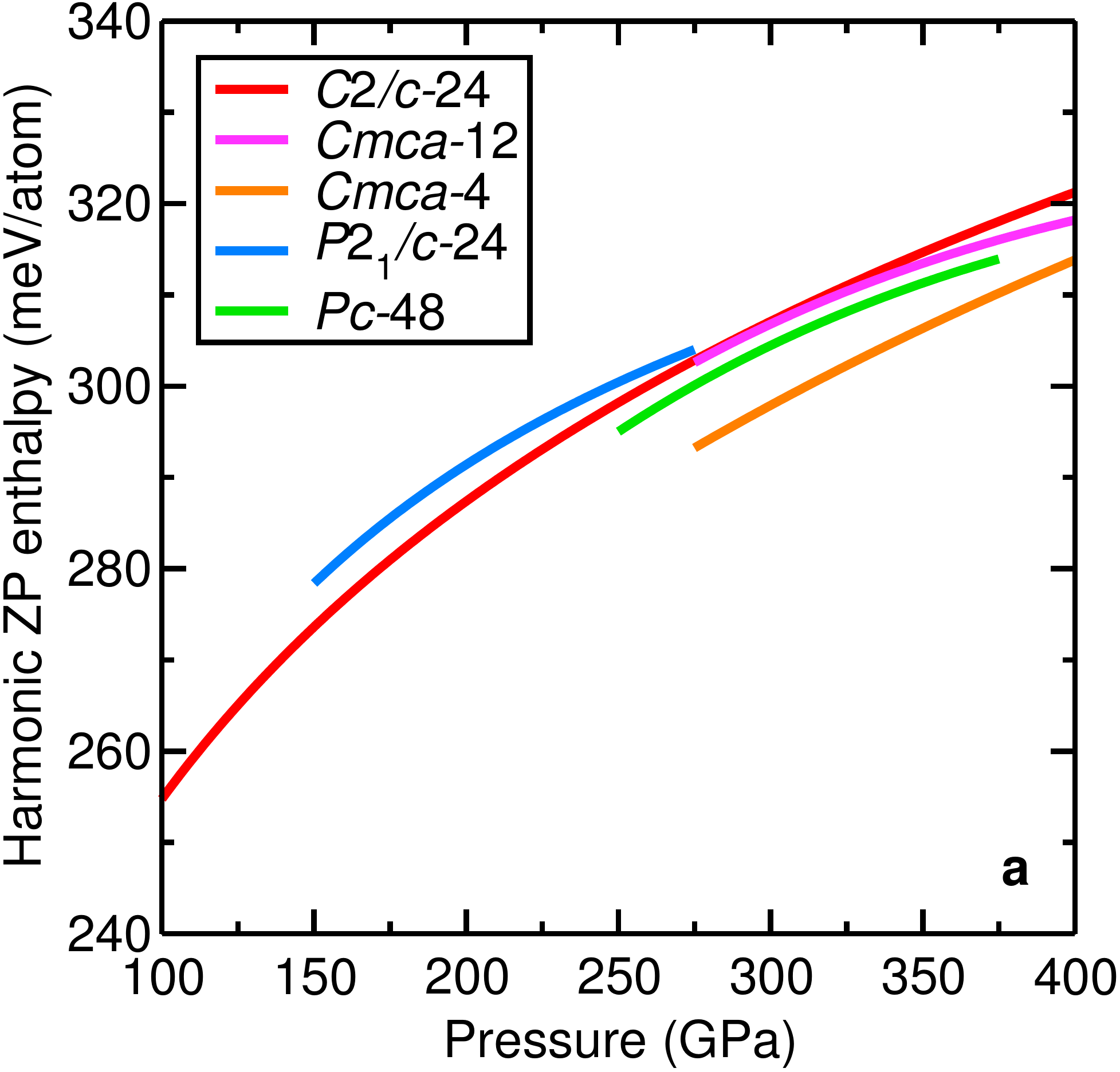}
&
   \includegraphics[width=0.3\textwidth]{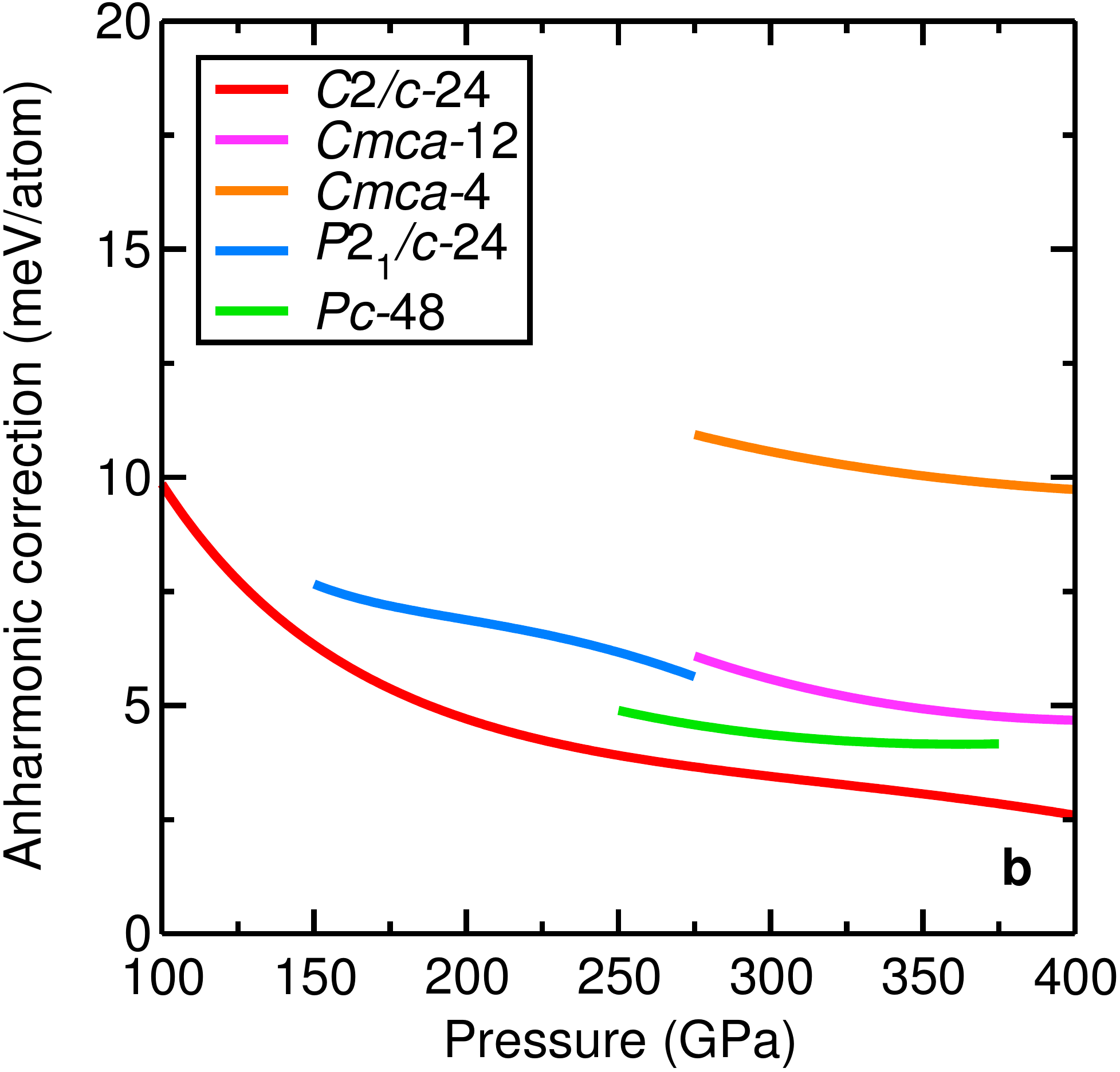}
\\
   \includegraphics[width=0.3\textwidth]{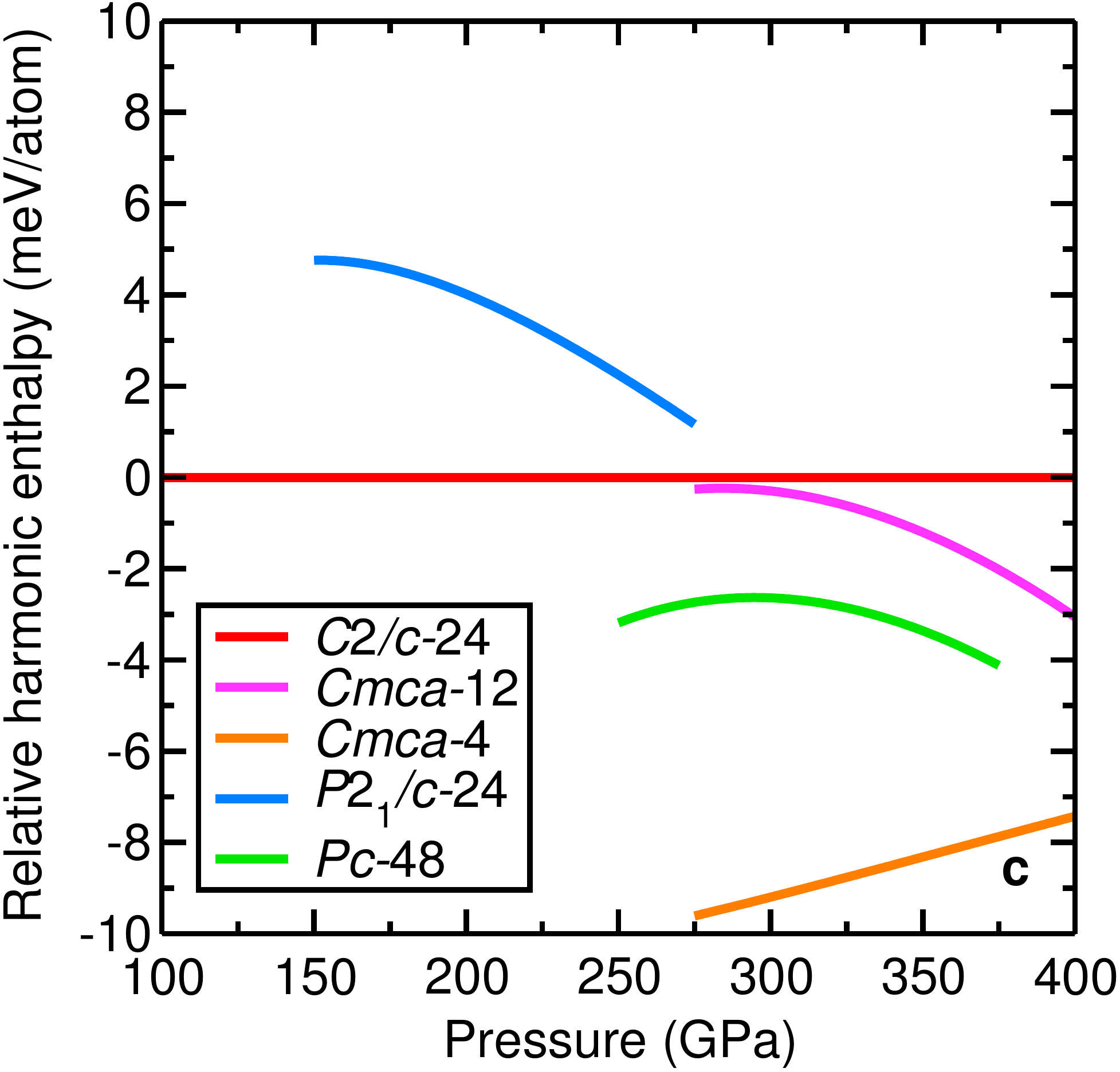}
&
   \includegraphics[width=0.3\textwidth]{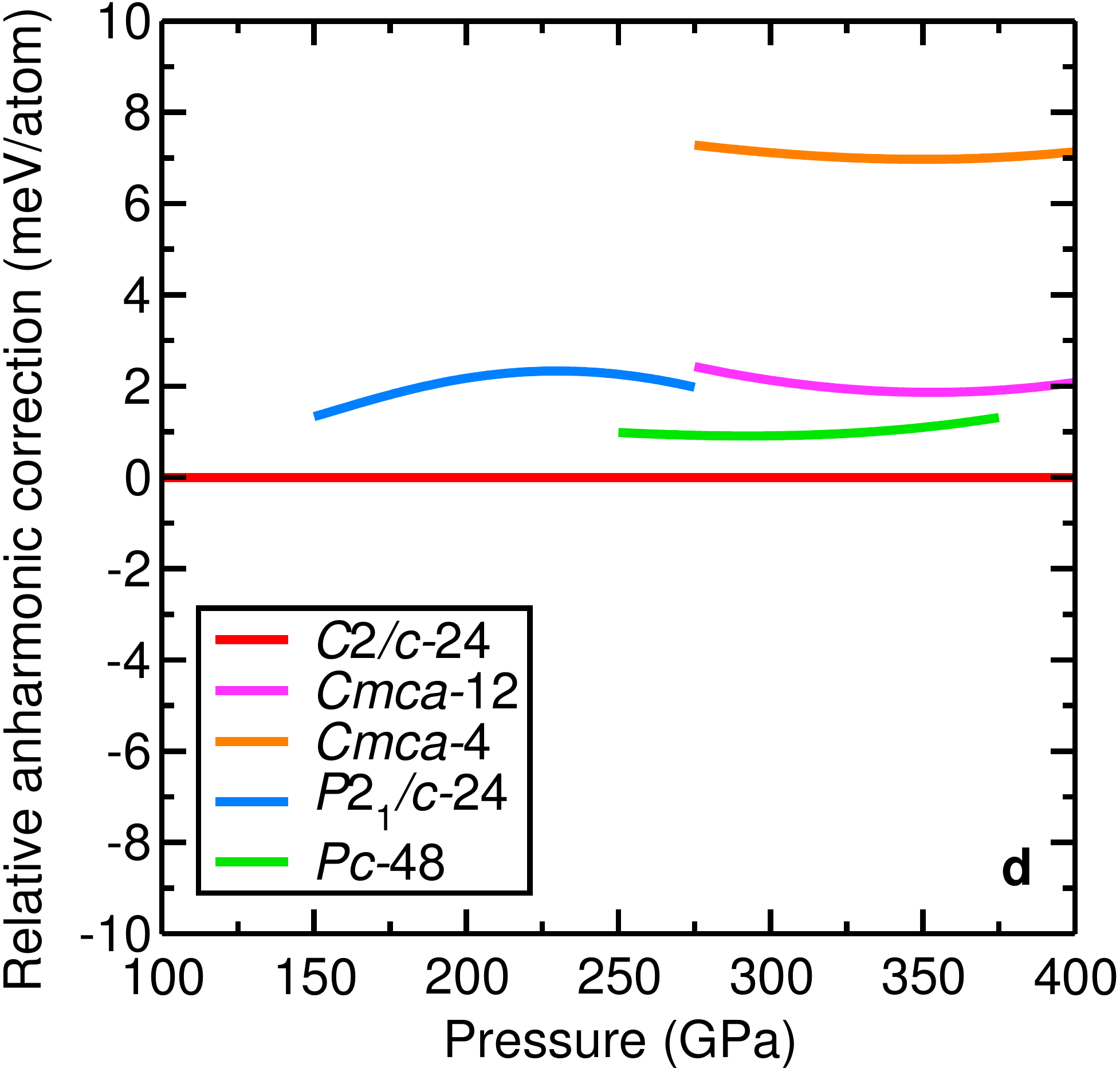}
\\
\end{tabular}
  \caption{\textbf{DFT-PBE vibrational contributions to the enthalpies of the H 
    structures.}
    (\textbf{a}) Harmonic ZP contributions to enthalpies, (\textbf{b}) 
    anharmonic ZP corrections to enthalpies, (\textbf{c}) harmonic ZP 
    enthalpies relative to \textit{C}2/\textit{c}-24, and (\textbf{d}) 
    anharmonic ZP corrections relative to \textit{C}2/\textit{c}-24.
    \textit{P}2$_{\text{1}}$/\textit{c}-24 is destabilised by both harmonic 
    vibrations and anharmonic corrections, relative to 
    \textit{C}2/\textit{c}-24.
    \textit{Cmca}-12, \textit{Cmca}-4, and \textit{Pc}-48 are all stabilised by
    harmonic vibrations but destabilised by anharmonic corrections, relative to
    \textit{C}2/\textit{c}-24.
  \label{fig:H_vibrations}}
\end{figure}

\paragraph{Vibrational results}
The harmonic zero-point (ZP) contributions to the enthalpies of the H phases 
increase sublinearly with pressure, as shown in Fig.\ \ref{fig:H_vibrations}(a),
while the anharmonic corrections tend to decrease with pressure; see 
Fig.\ \ref{fig:H_vibrations}(b). 
The harmonic ZP enthalpies are roughly thirty times larger than the anharmonic 
corrections. 
However, the differences between the harmonic ZP energies of the five phases 
considered at fixed pressure are similar in magnitude to the differences 
between the anharmonic corrections, both being about 10 meV per atom, as shown 
in Figs.\ \ref{fig:H_vibrations}(c) and
\ref{fig:H_vibrations}(d).
This demonstrates that the variations in the anharmonic vibrational corrections 
are as important as those of the harmonic contributions to the enthalpies in 
determining the relative stabilities of phases in this system.

\begin{figure}
  \includegraphics[width=0.225\textwidth]{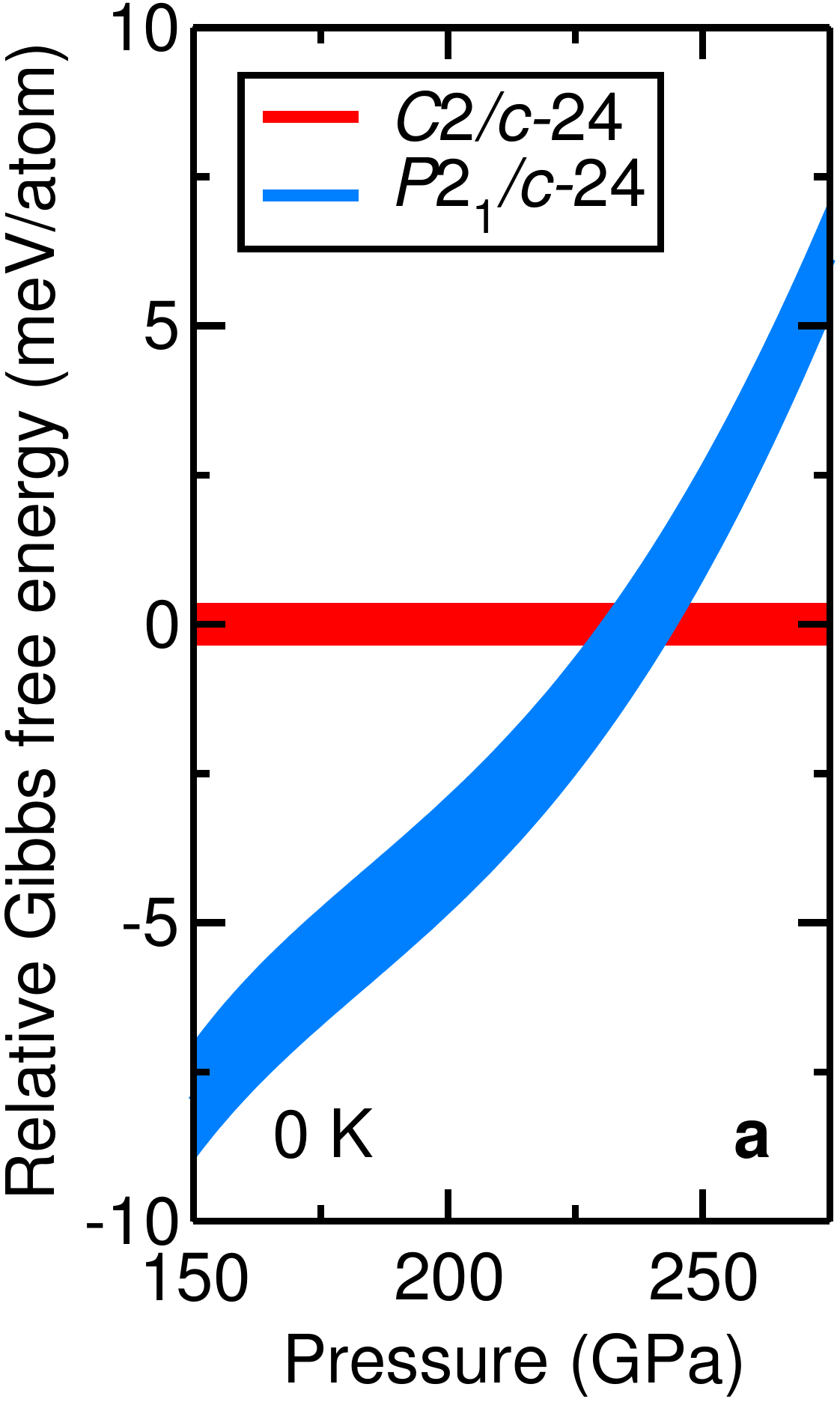}
  \includegraphics[width=0.225\textwidth]{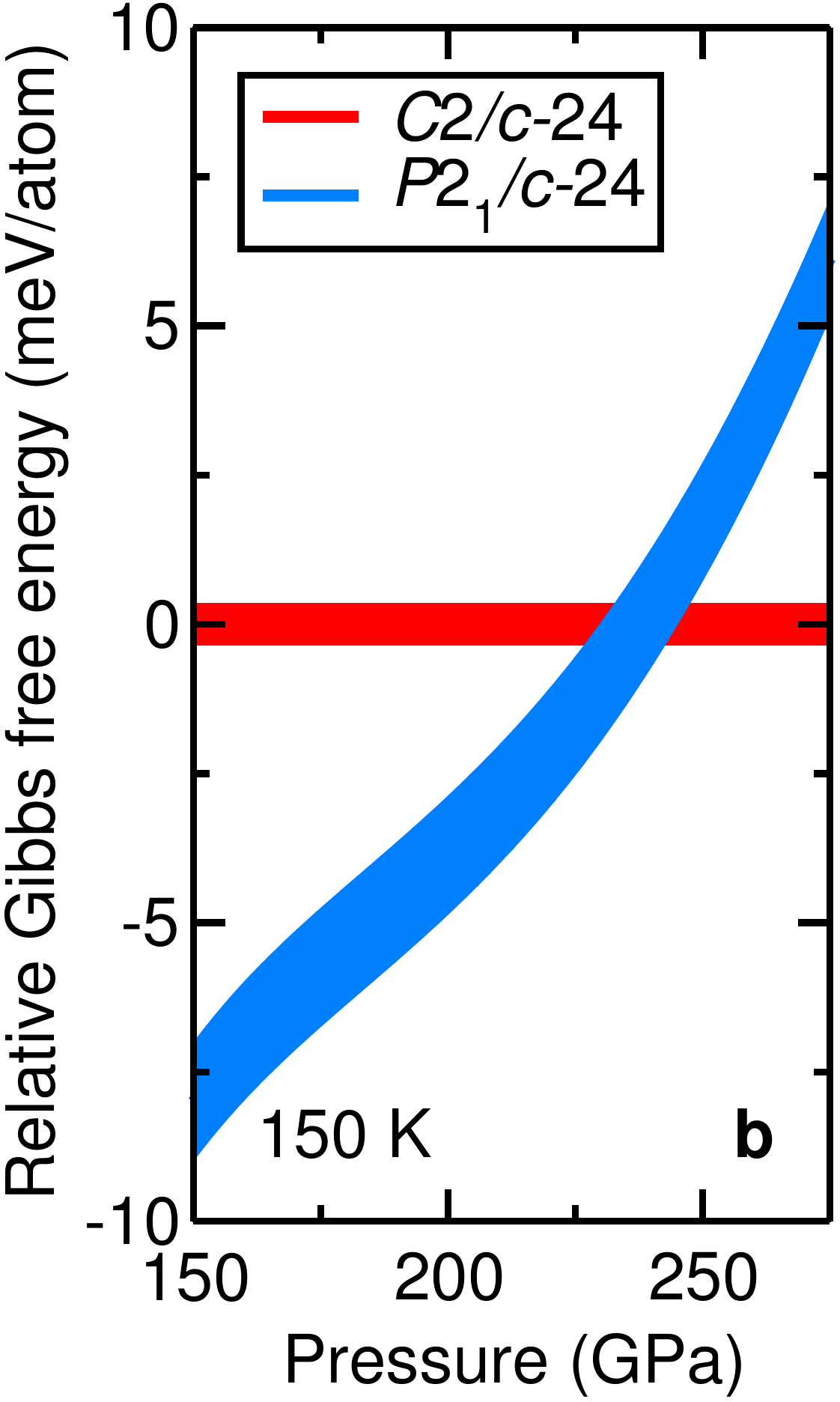}
  \includegraphics[width=0.225\textwidth]{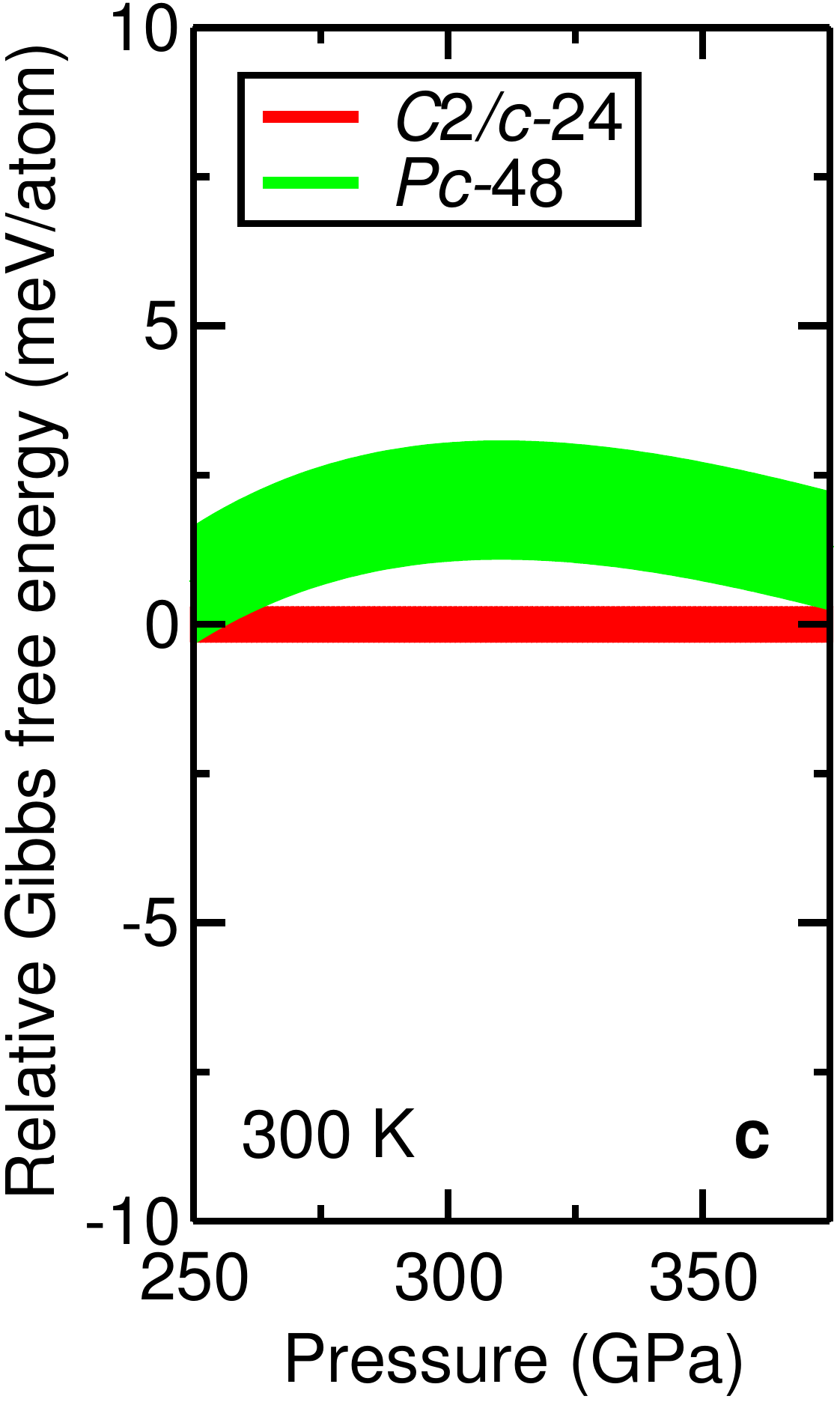}
  \includegraphics[width=0.225\textwidth]{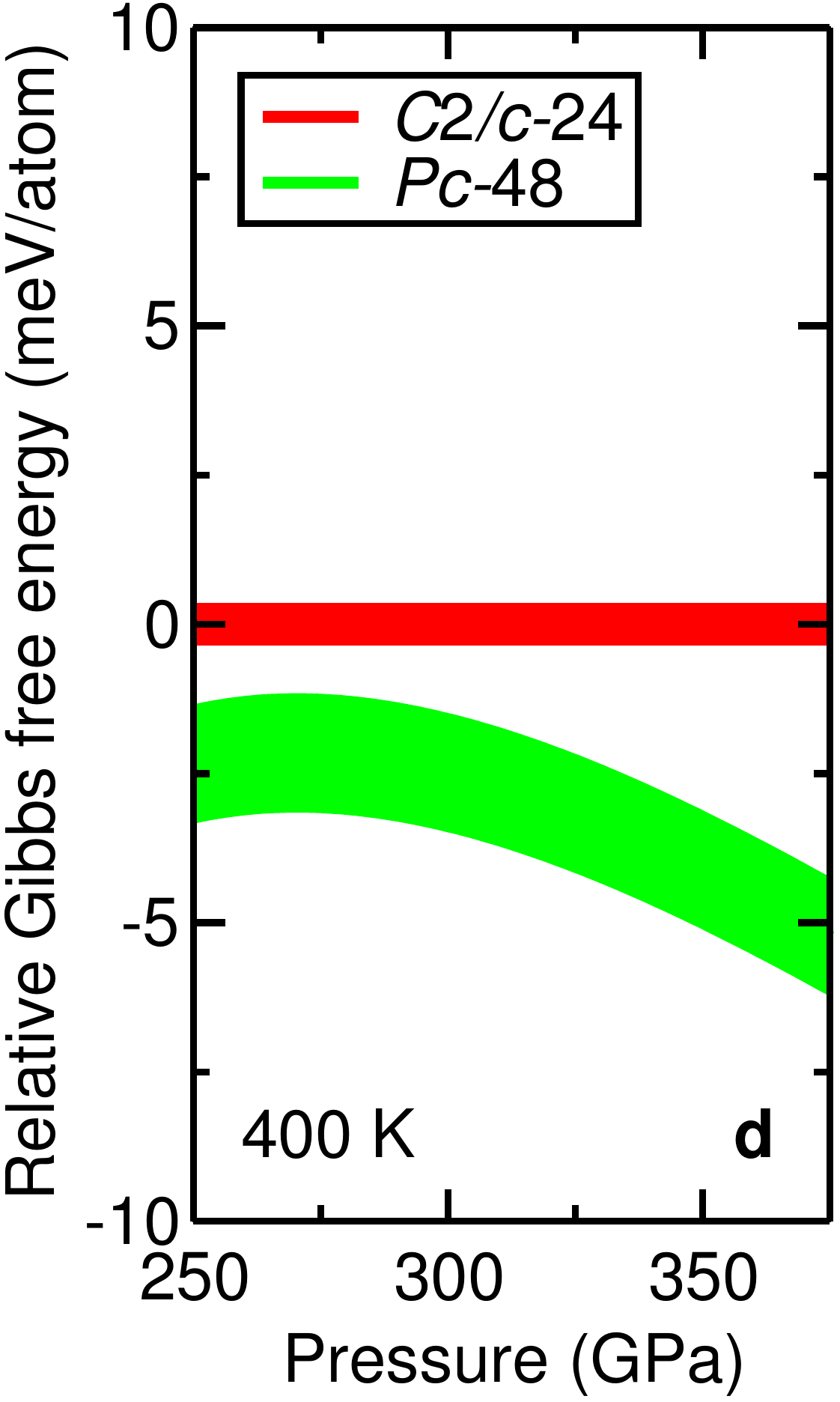}
  \caption{\textbf{Relative Gibbs free energies of the different H structures.} 
    (\textbf{a}) 0 K, (\textbf{b}) 150 K, (\textbf{c}) 300 K, and (\textbf{d}) 
    400 K\@.
    The Gibbs free energies were calculated using static-lattice DMC 
    calculations together with DFT-PBE harmonic and anharmonic vibrational
    calculations.  The transition from \textit{P}2$_{\text{1}}$/\textit{c}-24 to 
    \textit{C}2/\textit{c}-24 occurs at around $235 \pm 10$ GPa between 0 K
    and 150 K\@. \textit{Pc}-48 is stabilised by temperature with respect to 
    \textit{C}2/\textit{c}-24.  The complete set of relative enthalpies is
    shown in Supplementary Fig.\ 2.
  \label{fig:free_energy_vs_pressure_0K_300K_400K}}
\end{figure}

\begin{figure}
  \includegraphics[clip,width=0.45\textwidth]{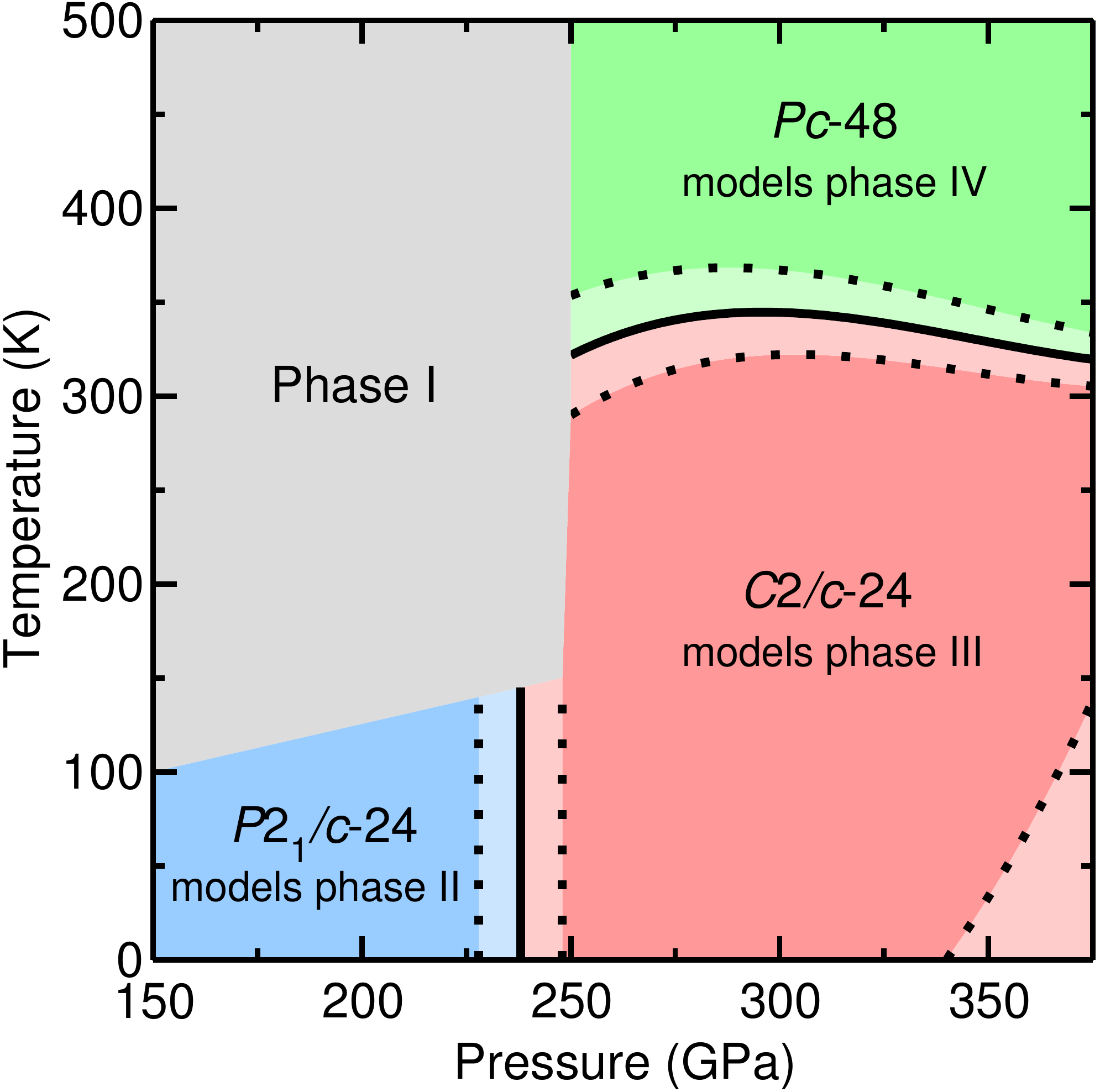}
  \caption{\textbf{Theoretical temperature-pressure phase diagram for H\@.}
    The solid black lines show the phase transitions calculated in this work,
    i.e., the set of points at which the relative Gibbs free energy of two
    phases is zero.  The dotted lines show the set of points at which the
    relative Gibbs free energy is one error bar from zero, and hence indicate
    the uncertainty in the phase boundaries.  At pressures in excess of
    350--375 GPa the Gibbs free energies of the \textit{C}2/\textit{c}-24 and
    \textit{Pc}-48 structures are within error bars of each other.  The grey
    region indicates the temperature-pressure conditions under which phase I
    is found to exist in experiments.
  \label{fig:phase_diagram_hydrogen}}
\end{figure}

\paragraph{Structural phase transitions}
Figure \ref{fig:free_energy_vs_pressure_0K_300K_400K} shows the two structural 
phase transitions that we have determined in this work, and our theoretical 
temperature-pressure phase diagram for solid molecular H is shown in Fig.\ 
\ref{fig:phase_diagram_hydrogen}. 
At 0 K, we find a transition from \textit{P}2$_{\text{1}}$/\textit{c}-24 to 
\textit{C}2/\textit{c}-24 at around $235 \pm 10$ GPa. 
The corresponding transition pressure for D is 13 GPa higher. (Note that the difference
between H and D is purely due to the DFT vibrational free energy and hence the
difference in transition pressures between H and D is relatively precise.)
Our transition pressure is around 75 GPa greater than those observed
experimentally for the transition between phases II and III, but the 13 GPa
difference between the transition pressures for H and D agrees well with the
experimentally measured value\cite{Goncharov_H_D_200GPa_2011}.  We note that
the theoretical transition pressures between H and D would only differ by
around 6 GPa without the inclusion of anharmonic effects.

As shown in Fig.\ \ref{fig:free_energy_vs_pressure_0K_300K_400K}, we also find 
a temperature-driven transition from \textit{C}2/\textit{c}-24 to 
\textit{Pc}-48 at pressures above 250 GPa and temperatures above 300 K, in good 
agreement with the experimentally observed transition between phases III and 
IV\@.
In Fig.\ \ref{fig:free_energy_vs_pressure_0K_300K_400K} we show the relative
free energies of \textit{C}2/\textit{c}-24 and \textit{Pc}-48 at 300 and 
400 K\@. 
At the lower temperature, \textit{C}2/\textit{c}-24 is marginally more stable, 
but at 400 K, \textit{Pc}-48 has clearly become the more stable structure.
The variation in the transition temperature with pressure is
  smaller than the uncertainty in that quantity, and so we report
  the \textit{C}2/\textit{c}-24--\textit{Pc}-48 transition temperature as $320
  \pm 20$ K\@.

\section*{Discussion}

Our theoretical H phase diagram is in reasonable quantitative agreement with
experiment, indicating that the \textit{P}2$_{\text{1}}$/\textit{c}-24, 
\textit{C}2/\textit{c}-24, and \textit{Pc}-48 structures provide satisfactory 
models for phases II, III, and IV\@. 
These model structures reproduce the experimental Raman and IR spectra quite 
well.
However, there is a significant disagreement of about 75 GPa between the
experimental and theoretical phase II--III transition pressure at 0 K\@.
There are several possible reasons for our substantially larger phase II--III
transition pressure. 
Firstly, the actual structure of phase III may be more stable than the 
\textit{C}2/\textit{c}-24 model structure. 
However, \textit{C}2/\textit{c}-24 is the most stable
non-metallic structure found in DFT
searches over a wide range of pressures, and it is compatible with the Raman
and IR spectra of the observed phase III\@.
If a significantly more stable structure
than \textit{C}2/\textit{c}-24 were to be found for phase III, the excellent description of the transition from
phase III to IV with increasing temperature obtained with our
calculated data would be spoilt.  
Another possible explanation for the discrepancy with experiment regarding the 
phase II--III transition pressure could be that we have neglected a significant 
contribution to the energy of \textit{P}2$_{\text{1}}$/\textit{c}-24 (our model 
for phase II)\@. 
In particular, our calculations do not account for nuclear exchange effects,
which are known to have a significant effect on the phase I--II transition 
pressure.\cite{Mazin_1997}
However, reliable estimates of the size of nuclear exchange effects in solid H
at high pressure are not currently available.  Furthermore,
  nuclear exchange effects are expected to be much smaller in D than H,
  because deuterons are bosons, whereas protons are fermions, and
  each deuteron has twice the mass of a proton.  This suggests that nuclear
  exchange effects cannot be entirely responsible for the discrepancy
  in the phase II--III transition pressure in both H and D\@.
Our analysis of different finite-size corrections in Supplementary 
Note 2 (see also the accompanying Supplementary Figs.\ 3 and 4) indicates that
finite-size effects in our relative enthalpies are
well-controlled, but it is always possible that finite-size effects may be
larger than anticipated.  Finally, the fixed-node approximation is an
uncontrolled source of error in our DMC calculations and, although fixed-node
errors should largely cancel when relative energies are calculated, it cannot
be ruled out that fixed-node errors may be larger in one phase than another.

The results we obtain by combining our DMC static-lattice energies and
harmonic and anharmonic vibrational
energies resolve a discrepancy between DFT and experiment
for the transition between phases III and IV\@.  The \textit{Pc}-48
structure was proposed as a candidate for phase IV in Refs.\
\onlinecite{Pickard_IV_2012} and\,
\onlinecite{Pickard_IV_2012_Erratum} because its Raman spectrum agrees
well with the experimental one and because its weakly bonded layers
lead to soft vibrational modes that thermally stabilise it.  However,
DFT static-lattice calculations together with the harmonic
approximation for nuclear motion (used in Refs.\
\onlinecite{Pickard_IV_2012} and\,
\onlinecite{Pickard_IV_2012_Erratum}) predict the \textit{Cmca}-4
structure to be energetically favoured at all temperatures in the
relevant pressure range.
(Note that \textit{Cmca}-4 is stabilised
  significantly by harmonic ZP energy; at the static-lattice DFT level it is not
  competitive, as shown in Fig.\ \ref{fig:DMC_vs_DFT}.)
The metallic nature of
\textit{Cmca}-4 contradicts experiment, in which insulating structures
containing strong molecular bonds are found up to pressures in excess
of 300 GPa.
The phase diagram predicted by DFT is shown in Supplementary Fig.\ 5.
The use of static-lattice DMC
energies and anharmonic vibrational energies destabilises
\textit{Cmca}-4, and we find that it is thermodynamically unstable
over the entire pressure and temperature range considered here.
Our results establish that DFT does not provide even a
  qualitatively correct description of the phase behaviour of hydrogen.
We also find that \textit{Cmca}-12 is unstable at the pressures and
temperatures studied in this work.  We have found an important
discrepancy between our calculated phase II--III transition pressure and experiment, which is
currently unresolved, although we have described possible physical reasons for
the disagreement.  Our calculations demonstrate that anharmonic
vibrational effects are crucial for determining the relative
stabilities of the phases.

\section*{Methods}

\paragraph{Quantum Monte Carlo calculations}
The DMC method\cite{Ceperley_1980,Foulkes_2001} is capable of delivering much 
higher accuracy than DFT, and the scaling of the computational cost with system 
size enables the simulation of the hundreds of atoms required for accurate 
calculations.
We have used the DMC method to calculate static-lattice energies using H
structures relaxed within DFT-PBE at a given external pressure. 
In DMC, the ground-state component of a trial wave function is projected out by
simulating the Schr\"{o}dinger equation in imaginary time, subject to the
constraint that the nodal surface of the wave function is fixed to be that of
the trial wave function\cite{Ceperley_1980,Foulkes_2001}. 
We used Slater-Jastrow wave functions as implemented in the {\sc casino}
code\cite{Needs_2010_CASINO}. Full technical details of our calculations can be
found in Supplementary Note 2.
The single-particle orbitals were obtained from the {\sc castep} code\cite{Clark_2005} using the PBE exchange-correlation functional. 
The nuclei were represented by bare Coulomb potentials and appropriate cusp 
corrections were applied to the orbitals. 
We used a flexible Jastrow factor\cite{Drummond_2004} whose parameters were 
optimised using variational Monte Carlo (VMC)\cite{Umrigar_2007}. 
VMC and DMC simulations were performed using 96 and 768 atoms, and the results 
were extrapolated to infinite system size. 
Using the resources of the Oak Ridge Leadership Computing Facility, we achieved
statistical error bars of less than 0.3 meV per atom in all our DMC 
calculations.

\paragraph{Anharmonic vibrational calculations}
We have calculated harmonic vibrational free energies by using the 
finite-displacement method to construct the matrix of force constants and 
diagonalising the corresponding dynamical matrices over a fine vibrational 
Brillouin-zone grid, as described in Supplementary Note 3 (with
accompanying data presented in Supplementary Figs.\ 6 and 7).
We determined anharmonic corrections to the harmonic free energies using a 
vibrational self-consistent field method\cite{Monserrat_VSCF_2013,
Monserrat_VSCF_helium_2014,Azadi_H_QMC_VSCF_2014}, sampling the low-energy part 
of the DFT-PBE Born-Oppenheimer (BO) energy surface along harmonic normal modes 
to large amplitudes. 
The resulting anharmonic Schr\"{o}dinger equation for the nuclear motion was 
solved by expanding the wave function in a basis of simple harmonic oscillator
eigenstates. 
Thermal occupation of excited states allowed us to calculate free energies at 
arbitrary temperatures. 
The vibrational free energy differences between the structures were converged to 
better than 1 meV per atom. 
Our approach does not describe possible melting.

\bibliography{hydrogen}

\begin{addendum}
 \item We thank Dominik Jochym for help with the implementation of the BLYP
   density functional. 
   Financial support was provided by the Engineering and Physical Sciences 
   Research Council (EPSRC), U.K\@.
   This research used resources of the Oak Ridge Leadership Computing Facility 
   at the Oak Ridge National Laboratory, which is supported by the Office of 
   Science of the U.S.\ Department of Energy under Contract No.\ 
   DE-AC05-00OR22725. 
   Additional calculations were performed on the Cambridge High Performance 
   Computing Service facility Darwin and the N8 high-performance computing 
   facility provided and funded by the N8 consortium and EPSRC (Grant No.\
   EP/K000225/1).
 \item[Author contributions] R.\,J.\,N.\ conceived and managed the project, 
   C.\,J.\,P.\ implemented the BLYP functional in the {\sc castep} code for our 
   work, P.\,L.\,R.\ optimised the trial wave functions for our DMC 
   calculations, J.\,H.\,L.-W.\ carried out the DFT calculations and calculated
   the DFT and DMC equations of state, B.\,M.\ carried out the anharmonic 
   vibrational calculations, and N.\,D.\,D.\ performed the DMC calculations.
   All authors contributed to the manuscript.
 \item[Competing Interests] We declare competing interests.
 \item[Correspondence] Correspondence and requests for materials should be 
   addressed to R.\,J.\,N.\ (email: rn11@cam.ac.uk).
\item[Supplementary Information] Supplementary information accompanies this
   paper.
\end{addendum}

\end{document}